\documentclass[a4paper,11pt]{article}
\pdfoutput=1

\usepackage{amssymb,mathrsfs,amsmath}
\usepackage{a4wide}
\usepackage{color,xcolor}
\usepackage{slashed,soul}
\usepackage{graphicx}
\usepackage{caption}
\usepackage{subcaption}
\usepackage{amsfonts}
\usepackage{comment}
\usepackage{lscape}
\def\linkcolor{cyan!70!black}

\usepackage[
colorlinks=true
,urlcolor=\linkcolor
,anchorcolor=\linkcolor
,citecolor=\linkcolor
,filecolor=\linkcolor
,linkcolor=\linkcolor
,menucolor=\linkcolor
,linktocpage=true
,pdfproducer=medialab
,pdfa=true
]{hyperref}
\usepackage{amsthm}
\usepackage{booktabs}
\usepackage{array}
\usepackage{rotating}
\usepackage[numbers, sort&compress]{natbib}
\usepackage{multirow}
\usepackage{float}
\usepackage[utf8]{inputenc}
\usepackage[T1]{fontenc}
\usepackage{appendix}
\usepackage{orcidlink}

\renewcommand{\baselinestretch}{1.2}

\let\OLDthebibliography\thebibliography
\renewcommand\thebibliography[1]{
  \OLDthebibliography{#1}
  \setlength{\parskip}{0pt}
  \setlength{\itemsep}{0pt plus 0.3ex}
}

\providecommand{\abs}[1]{\lvert#1\rvert}

\allowdisplaybreaks

\begin{document}

\vspace{1cm}

\vspace{0.8truecm}

\begin{center}
\renewcommand{\baselinestretch}{1.8}\normalsize
\boldmath
{\LARGE\textbf{ Lepton Flavor-Violating Higgs Decays Mediated by Ultralight Gauge Boson
}}
\unboldmath
\end{center}

\vspace{0.4truecm}

\renewcommand*{\thefootnote}{\fnsymbol{footnote}}

\begin{center}
{\bf 
 Marcela Mar\'in$\,^1$\footnote{\href{mailto:marcemarino8a@cuautitlan.unam.mx}{marcemarino8a@cuautitlan.unam.mx}}\orcidlink{0000-0001-8520-6582},
 R. Gait\'an$\,^1$\orcidlink{0000-0001-7212-8722}, 
 R. Martinez$\,^2$\orcidlink{0000-0002-3376-6307}
\vspace{0.5truecm}}

{\footnotesize
$^1$ {\sl Departamento de F\'isica, FES-Cuautitl\'an, UNAM, C.P. 54770, Estado de M\'exico, M\'exico.\vspace{0.15truecm}}

$^2$ {\sl Departamento de F\'isica, Universidad Nacional de Colombia, K. 45 No. 26-85, Bogot\'a, Colombia. \vspace{0.15truecm}}
}
\end{center}

\renewcommand*{\thefootnote}{\arabic{footnote}}
\setcounter{footnote}{0}

\vspace{0.4cm}
\begin{abstract}
\noindent 
We present an analysis of the lepton-flavor violating decay of the Higgs boson mediated by an ultralight gauge boson, $\chi$. Our analysis matches a model generating the lepton flavor-violating interaction $\bar{\ell}_i\ell_j\chi$ at tree level with an effective field theory, safeguarding a physical massless $\chi$-boson limit of the observables. By utilizing the upper bounds on $H\to\ell_i \bar\ell_j$ decays from CMS and ATLAS Collaborations, we establish an indirect upper limit on the nonstandard decay $H\to\ell_i \bar\ell_j\chi$. The analysis encompasses various observables such as the lepton energy spectrum, Dalitz plot distribution, and Lepton Charge and Forward-Backward Asymmetries.
\end{abstract}

\section{Introduction}
The discovery of the Higgs boson in 2012 by the ATLAS and CMS Collaborations at the Large Hadron Collider (LHC), with a mass of approximately $125~{\rm GeV}$ \cite{ATLAS:2012yve, CMS:2012qbp, CMS:2013btf}, marked a significant milestone in our understanding of the Standard Model (SM). Since then, precise measurements of its properties, including decay branching fractions, have consistently aligned with the expectations of the SM \cite{ATLAS:2022vkf, CMS:2022dwd, ATLAS:2015egz, CMS:2014fzn, CMS:2012vby, ATLAS:2013xga, CMS:2014nkk, CMS:2017dib}. However, amidst this conformity lies an intriguing frontier: exploring decay channels that deviate from the SM predictions. While extensive efforts are underway to measure the decay rates within SM-favored channels precisely, there is a corresponding need for commensurate attention to be directed toward those subdominant or absent channels within the SM. Especially intriguing are scenarios in which new physics phenomena may emerge, potentially surpassing the expectations set forth by the SM.

The SM assumption that left-handed neutrinos are massless implies conserving the lepton family number. However, neutrino oscillation experiments have unequivocally demonstrated that lepton flavor conservation is not a fundamental symmetry of nature \cite{Super-Kamiokande:1998kpq, SNO:2002tuh}. In the minimal extension of the SM where neutrinos have nonvanishing masses, charged Lepton Flavor Violation (cLFV) is made possible through neutrino oscillation. Even so, cLFV processes are heavily suppressed by the GIM mechanism, making them unobservable in current experiments. Nevertheless, the potential detection of cLFV would provide an enticing avenue for exploring physics Beyond the Standard Model (BSM).

Lepton-flavor violating (LFV) decays \(H\rightarrow\mu e\)\footnote{In this manuscript, the notation $H \to \mu e$ refers to $H \to \mu \bar{e}$ and 
$H \to \bar{\mu} e$. The same convention applies to the other cLFV modes.}, \(H\rightarrow\tau e\), or \(H\rightarrow\tau\mu\) are strictly forbidden within the confines of the SM framework and effectively suppressed in its minimal extension \cite{Arganda:2004bz}.  However,  they find theoretical justification in BSM theories, where LFV interaction can arise through off-diagonal LFV Yukawa couplings, which facilitate the interaction of the Higgs boson with leptons of differing flavors. These processes are primarily induced via virtual boson exchange mechanisms \cite{Diaz-Cruz:1999sns, Arhrib:2012ax, Goudelis:2011un, Pilaftsis:1992st, Ishimori:2010au}, often occurring within extended Higgs sectors, or involving several scalar doublets \cite{Bjorken:1977vt, Branco:2011iw, Agashe:2009di, Azatov:2009na, Lami:2016mjf, Han:2000jz, Brignole:2003iv}.  LHC Collaborations have undertaken experimental efforts to probe Higgs decays \cite{CMS:2023pte, CMS:2021rsq, ATLAS:2019old}. Specifically, the ATLAS and CMS Collaborations have conducted extensive measurements for the SM decay modes \(H\rightarrow\tau\bar\tau\) and \(H\rightarrow\mu\bar\mu\), alongside establishing upper bounds for \(H\rightarrow e\bar e\) and all potentially cLFV channels of \(H\rightarrow\ell_i\bar\ell_j\) decays. The outcomes of these searches are detailed in Table \ref{Table:Br}.

\begin{table}[t!] 
\begin{center}
    \begin{tabular}{c c c|c c c}
    \hline
    \hline
    SM modes   & BR & &cLFV modes& BR ($95\%$CL)& \\ \hline
     $H\to\tau\bar\tau$   & $(6.0^{+0.8}_{-0.7})\%$ & \cite{ATLAS:2022vkf}&$H\to\tau\mu$ & $\leq 1.5\times10 ^{-3}$& \cite{CMS:2021rsq}\\
     $H\to \mu \bar\mu$ & $(2.6\pm 1.3)\times10^{-4}$ & \cite{ATLAS:2022vkf} & $H\to\tau e$& $\leq 2.2\times10^{-3}$ & \cite{CMS:2021rsq}\\
     $H\to e\bar e$&  $\leq 3.6\times10^{-4}$ ($95\%$CL) & \cite{ATLAS:2019old} & $H\to \mu e$& $\leq 6.1\times10^{-5}$ & \cite{ATLAS:2019old}\\   
     \hline
\hline
    \end{tabular}
\end{center}
\caption{Experimental measured and upper bounds on the branching fraction of $H\to\ell_i \bar\ell_j$.}
\label{Table:Br}
\end{table}

This paper presents the phenomenological implications of cLFV Higgs decays, \(H\to\ell_i \bar\ell_j\), mediated by an ultralight gauge boson, $\chi$, where the cLFV interaction with this boson arising at tree level. Additionally, we delve into the scenario where \(\chi\) is on-shell establishing an upper bound for these decays. This paper is structured as follows: Section \ref{EFL} provides an overview of the model inducing the cLFV interactions and its correspondence with an effective Lagrangian. Section \ref{LFV-Higg-D} discusses the analytical outcomes of cLFV Higgs decays mediated by \(\chi\), accompanied by phenomenological analyses of these processes. We analyze scenarios with the gauge boson off-shell in Section \ref{Off-S} and on-shell in Section \ref{On-S}. Finally, Section \ref{Conclusions} presents our conclusions.

\section{Effective Lagrangian description}\label{EFL}
In the letter \cite{Ibarra:2021xyk},  one of the authors proposes two models enabling LFV interactions via a gauge boson, denoted as $\chi$.  Their focus lies on an ultralight gauge boson, which ensures the finiteness of the observables in the massless $\chi$-boson limit. Although the initial paper confines the models to two generations, expanding them to encompass three generations is straightforward \cite{Marin:2022wfk}. We are especially intrigued by the tree-level model that induces cLFV transitions at the tree level.

In the tree-level model, a $U(1)_\chi$ gauge symmetry was introduced alongside complex scalar fields, $\phi_{ij}$, $i\,,j=1\,,2$. The model postulated that leptons possess generation-dependent charges under $U(1)_\chi$. Additionally, the tree-level model considered that doublet scalars acquire a vacuum expectation value, $\langle \phi_{jk}\rangle =v_{jk}$, where the hypercharge $Y_{jk}=1/2$, and $q_{\phi_{jk}}=q_{L_j}-q_{e_k}$. Table \ref{tab:tree_level} summarizes the particle content and the corresponding spins and charges under the symmetries. Here, $L_i=(\nu_{L_i}, e_{L_i})$ and $e_{R_i}$, where $i=1\,,2$, represent the standard model $SU(2)_L$ lepton doublets and singlets, respectively.

\begin{table}[t!]
\begin{center}
\begin{tabular}{ |c| c c c c | c c c c|}
  \hline
  & $L_1$ & $L_2$ & $e_{R_1}$ & $e_{R_2}$ & $\phi_{11}$& $\phi_{12}$& $\phi_{21}$& $\phi_{22}$ \\ 
  \hline
  spin & 1/2 & 1/2 & 1/2 & 1/2 & 0 & 0 & 0 & 0\\
 $SU(2)_L$ & 2 & 2 & 1  & 1 & 2 & 2 & 2 & 2 \\
 $U(1)_Y$ & $-1/2$ & $-1/2$ & $-1$ & $-1$ & $Y_{11}$ &  $Y_{12}$ & $Y_{21}$ &  $Y_{22}$\\
 $U(1)_\chi $ & $q_{L_1}$ & $q_{L_2}$ & $q_{e_1}$ & $q_{e_2}$ & $q_{\phi_{11}}$& $q_{\phi_{12}}$& $q_{\phi_{21}}$& $q_{\phi_{22}}$\\
 \hline
\end{tabular}
\end{center}
\caption{Spins and charges under $SU(2)_L \times U(1)_Y \times U(1)_\chi$ model describing LFV transitions at tree level in the two generation case. All fields are assumed to be singlets under $SU(3)_C$.}
\label{tab:tree_level}
\end{table}

The kinetic and the Yukawa Lagrangian are formulated as follows:
\begin{align}
{\cal L}_{\rm kin}&=\sum_{j=1}^2 i( \overline L_j\slashed{D} L_j+ \overline e_{R_j} \slashed{D} e_{R_j})+\sum_{j,k=1}^2 (D_\mu \phi_{jk})^\dagger (D^\mu \phi_{jk} )\;,\nonumber\\
-{\cal L}_{\rm Yuk}&=\sum_{j,k=1}^2 y_{jk} \overline L_j \phi_{jk} e_{R_k}+{\rm h.c.}\,,
\end{align}
where $D_\mu$ denotes the covariant derivative, given by
\begin{align}
D_\mu&=\partial_\mu +i g W_\mu^a T_a+i g' Y B_\mu + i g_\chi q \chi_\mu~~{\rm for~the }~SU(2)_L~{\rm doublets}\;, \nonumber \\
D_\mu&=\partial_\mu +i g' Y B_\mu + i g_\chi q \chi_\mu ~~{\rm for~the}~SU(2)_L~{\rm singlets}\;,
\label{eq:L_kin_tree_level}
\end{align}
being $g$, $g'$ and $g_\chi$ the coupling constants of $SU(2)_L$, $U(1)_Y$ and $U(1)_\chi$, respectively. 

The LFV vertex arises from recasting the kinetic Lagrangian  in terms of the mass eigenstates
\begin{align}
-{\cal L}\supset \overline{e_R} i g_{e \mu}^{RR} \gamma^\rho \chi_\rho  \mu_R +\overline{e_L} i g_{e \mu}^{LL} \gamma^\rho \chi_\rho \mu_L +{\rm h.c.}\;,
\label{eq:LFV-tree-level-model}
\end{align}
with
\begin{align}
g_{e\mu}^{RR}&=g_\chi (q_{e_{1}}-q_{e_{2}})\sin\theta_R\cos\theta_R \;,\nonumber \\
g_{e\mu}^{LL}&=g_\chi (q_{L_{1}}-q_{L_{2}})\sin\theta_L\cos\theta_L\;,
\end{align}
where it can be deduced that the mixing angles $\theta_L\,,~\theta_R$ and the masses of the particles can be written as a function of $y_{jk}$, $v_{jk}$ and $q_{\phi_{jk}}$, see Appendix \ref{apx:masses}. Evidently, when the $U(1)_\chi$ charges are generation-independent, lepton flavor is exactly conserved within the model. Likewise, if the interaction eigenstates coincide with the mass eigenstates, tree-level flavor-changing interactions are absent (see Eq. (\ref{eq:mass_matrix})). This alignment of eigenstates can occur under specific conditions, such as particular configurations of the vacuum expectation values of the fields $\phi_{ij}$, for example, when $v_{ii} = 0$ but $v_{ij} \neq 0$. This mechanism holds not only for two generations but can also be extended to three generations \cite{Marin:2022wfk}.

We can match this model with an Effective Field Theory (EFT), which offers the advantage of greater generality and fewer constraints.  Within the framework of general EFT, the interaction responsible for lepton flavor violation can be described by the effective Lagrangian
\begin{equation}
    \mathcal{L}\supset F_1 \bar{e}\gamma_\alpha\chi^\alpha\mu+G_1 \bar{e}\gamma_\alpha\gamma_5\chi^\alpha\mu+{\rm h.c.}\,,
\end{equation}
 where compared to  the Lagrangian in Eq.~(\ref{eq:LFV-tree-level-model}), we match the effective couplings, $F_1$ and $G_1$, to correspond with the parameters of the model
 \begin{equation}
    F_1=\frac{1}{2}\left(g_{e\mu}^{RR}+g_{e\mu}^{LL}\right)\,, \quad G_1=\frac{1}{2}\left(g_{e\mu}^{RR}-g_{e\mu}^{LL}\right)\,.
    \label{eq:FF}
\end{equation}

For simplicity, we assume for the moment that the Yukawa couplings satisfy $y_{22}>>y_{11}>>y_{12},y_{21}$ and all $v_{jk}=v$, so  $q_{\phi_{jk}}=Q$. The relevant parameters of the tree-level model after the spontaneous breaking of the gauge symmetries are:
\begin{align}\label{eq:parameters}
    &m_{\mu}^2\simeq y_{22}^2v^2\,,\quad m_e^2\simeq y_{11}^2v^2\,\quad m_\chi^2\simeq 4g_{\chi}^2Q^2v^2\,,\nonumber\\
    &\sin{2\theta_L}\simeq-2\frac{y_{12}}{y_{22}}\,,\quad\sin{2\theta_R}\simeq-2\frac{y_{21}}{y_{22}}\,.
\end{align}

In this work, we focus on the scenario where \(\chi\) is (ultra)light, specifically \( m_\chi < m_\mu \) (with \( m_\chi \ll m_\mu \) in the ultralight case). The effective parameters can be constrained to ensure the correct muon mass \( m_\mu \sim 105 \) MeV by setting appropriate limits on the Yukawa couplings, \( U(1)_\chi \)-charges, and the gauge coupling \( g_\chi \), along with the vacuum expectation values of the scalar doublets \( \phi_{ij} \). To achieve a light \(\chi\)-boson, one may select a small \( g_\chi \), small \( q_{\phi_{ij}} \), or both, leading to an ultralight \(\chi\)-boson while preserving the observed lepton mass spectrum.

Based on the assumptions in Eqs.~(\ref{eq:parameters}), it can be found that $F_1$ and $G_1$ can be expressed as:
\begin{align}\label{eq:F1_G1}
   & F_1=\frac{1}{4Q}\left[\left(q_{e_2}-q_{e_1}\right)y_{21}+\left(q_{L_1}-q_{L_2}\right)y_{12}\right]\frac{m_\chi}{m_\mu}\equiv c^v\frac{m_\chi}{m_\mu}\,,\nonumber\\
   &G_1=\frac{1}{4Q}\left[\left(q_{e_2}-q_{e_1}\right)y_{21}-\left(q_{L_1}-q_{L_2}\right)y_{12}\right]]\frac{m_\chi}{m_\mu}\equiv c^a\frac{m_\chi}{m_\mu}\,.
\end{align}

Expanding to three generations, the model that induces LFV transitions at the tree level is included in the low-energy effective Lagrangian with monopole operators, as follows: 
\begin{equation} \mathcal{L}_{\rm eff}=f_{ij}\bar{\ell}_i\gamma^\alpha \chi_\alpha \ell_j + g_{ij} \bar{\ell}_i\gamma^\alpha \gamma_5 \chi_\alpha \ell_j  + {\rm h.c.} , \label{eq:Eff-Lag}
\end{equation}
where $\ell_i\, ,\ell_j=e\,,\mu\,,\tau$ and $f_{ij}\,,~g_{ij}$ are the effective couplings in the three generations case. Taking into account the matching with the parameters of the model in Eqs.~(\ref{eq:F1_G1}), we can simplify and define:
\begin{equation}\label{eq:f-g}
f_{ij}= c^v_{ij}\frac{m_\chi}{m_{\ell_i^j}} ~~{\rm and}~~ g_{ij}= c^a_{ij}\frac{m_\chi}{m_{\ell_i^j}}\,,  
\end{equation}
where $m_{\ell_i^j}$, represents the mass of the highest-generation lepton between $\ell_i$ and $\ell_j$, and $c^{v}_{ij}$ and $c^a_{ij}$ are dimensionless independent coefficients.

In \cite{Ibarra:2021xyk}, a model inducing LFV interaction at the one-loop level is also proposed. Although our study focuses on LFV interactions at the tree level, it is noteworthy that the one-loop level model introduces both monopole and dipole operators in an effective Lagrangian.

\section{LFV Higgs decays}\label{LFV-Higg-D}
In the Standard Model, the flavor-conserving interaction \( \bar{\ell}_i\ell_iH \) emerges, while our effective Lagrangian, as outlined in Eq.~(\ref{eq:Eff-Lag}), describes transitions for all lepton interactions, encompassing scenarios with flavor-conserving interactions  \( \ell_i = \ell_j \) and flavor-violating interactions \( \ell_i \neq \ell_j \). Using this effective Lagrangian, we can induce LFV processes \( H\to\ell_i \bar\ell_j \) at the loop level mediated by \( \chi \), as well as \( H\to\ell_i \bar\ell_j\chi \) at the tree level. The forthcoming two subsections will present the analytical and phenomenological outcomes of these decay processes.

\subsection[Higgs LFV decays with $\chi$ off-shell]{Higgs LFV decays with $\boldsymbol{\chi}$ off-shell}\label{Off-S}

\begin{figure}[t!]
\centering
\includegraphics[width=0.8\textwidth]{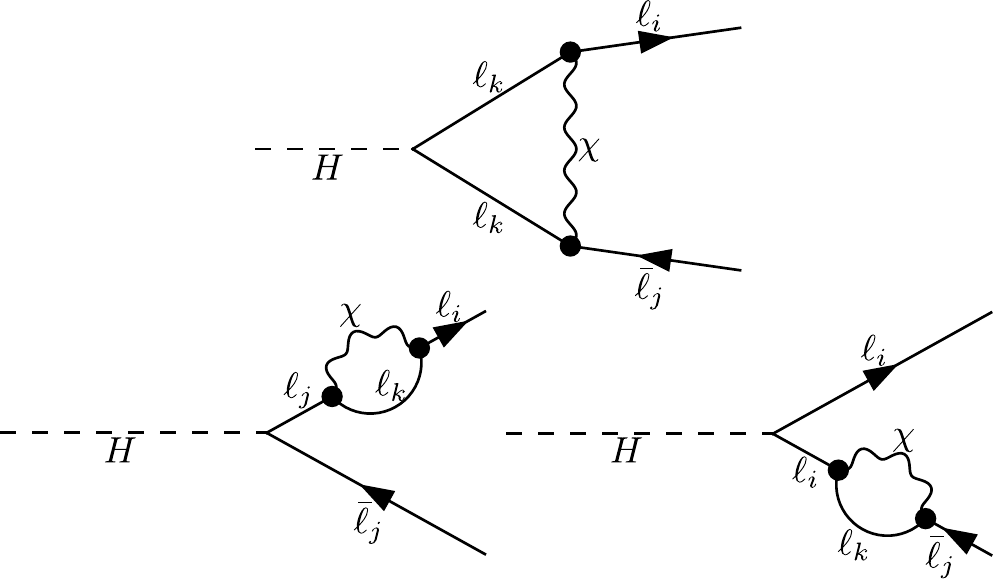} 
\caption{One-loop contribution to $H\to\ell_i\bar{\ell_j}$, with $\ell_i\,,\ell_j\neq\ell_k$ and $\ell_i\,,\ell_j=\ell_k$.}
\label{fig:hll-loop}
\end{figure}

Equation (\ref{eq:Eff-Lag}) describes the Lagrangian that induces one-loop level LFV decays of \( H\to\ell_i\bar{\ell}_j \) for all possible flavor combinations, including \( H\to\mu e \), \( \mu\tau \), and \( e\tau \), mediated by the \( \chi \)-boson, as depicted in Figure \ref{fig:hll-loop}. When the $\chi$-boson is ultralight, the triangle diagram contribution becomes predominant in \( H\to\ell_i\bar{\ell}_j\) processes, which justifies our focus on this particular contribution in the analysis.  The contribution from the triangle diagram to the branching ratio of \( H\to\ell_i\bar\ell_j \), with \( \ell_k \) into the loop, is given by:
\begin{align} \label{eq:BR-Hll} 
    {\rm BR}(H\to\ell_i\ell_j)=&\frac{\Gamma(H\to\ell_i\bar{\ell}_j)+\Gamma(H\to\bar{\ell}_i\ell_j)}{\Gamma_H}\\
    \simeq\frac{ m_\chi^4}{m_{\ell_{i}^{k}}^2 m_{\ell_{j}^{k}}^2}&\frac{M_H m_{\ell_k}^2}{4\pi~\Gamma_H~  v^2}
    \bigl[\abs{c_{jk}^v c_{ik}^a-c_{ik}^v c_{jk}^a}^2+\abs{c_{jk}^v c_{ik}^v-c_{ik}^a c_{jk}^a}^2\bigr]~\bigl\lvert \mathcal{F}_{{\rm ren}}(m_{\ell_i}\,,m_{\ell_j}\,,m_{\ell_k})\bigr\rvert^2\,,  \nonumber
\end{align}
where we have conveniently neglected the masses of the leptons in the kinematic expression, \(\Gamma_H\) represents the total Higgs decay width, and the loop function \(\mathcal{F}(m_{\ell_i}, m_{\ell_j}, m_{\ell_k})\) is described by
\begin{align}\label{Eq:Loop-Function}
 \mathcal{F}(m_{\ell_i}\,,m_{\ell_j}\,,m_{\ell_k})=&\frac{{\rm A_0}\left[m_{\ell_k}^2\right]}{m_\chi^2}-\left(\frac{M_H^2}{2m_\chi^2}-\frac{2m_{\ell_k}^2}{m_\chi^2}-3\right){\rm B_0}\left[M_H^2\,,m_{\ell_k}^2\,,m_{\ell_k}^2\right] -3\Bigl({\rm B_0}\left[m_{\ell_i}^2\,,m_{\ell_k}^2\,,m_\chi^2\right]\nonumber\\
 +{\rm B_0}&\left[m_{\ell_j}^2\,,m_{\ell_k}^2\,,m_\chi^2\right]-\left(M_H^2+m_\chi^2\right){\rm C_0}\left[M_H^2\,,m_{\ell_i}^2\,,m_{\ell_j}^2\,,m_{\ell_k}^2\,,m_{\ell_k}^2\,,m_\chi^2\right]\Bigr)+2 \,.
\end{align}
A$_0$, B$_0$, and C$_0$ denote the Passarino-Veltman integrals. The function \(\mathcal{F}(m_{\ell_i}, m_{\ell_j}, m_{\ell_k})\) depends on the masses involved in the process, including the Higgs mass, the lepton masses, and the \(\chi\)-boson mass. The loop function in Eq.~(\ref{Eq:Loop-Function}) exhibits UV divergences, and it can be expressed as the sum of a divergent and a finite (renormalized) part, denoted as $\mathcal{F}= \mathcal{F}_{\text{ren}} + \mathcal{F}_{\text{div}}$ where \(\mathcal{F}_{\text{ren}}\) refers to the renormalized contribution, and \(\mathcal{F}_{\text{div}}\) captures the divergent component (see Appendix \ref{apx:UV-Divergent} for details). It is important to note that the \(\chi\)-boson mass in the denominator of \(\mathcal{F}(m_{\ell_i}, m_{\ell_j}, m_{\ell_k})\) is canceled by the \(m_\chi^2\) term in the expression for \({\rm BR}(H\to\ell_i \ell_j)\) in Eq.~(\ref{eq:BR-Hll}), thus ensuring finitude in the massless limit of \(m_\chi\).

From Eq.~(\ref{eq:BR-Hll}), it is evident that the branching ratio \( {\rm BR}(H\to\ell_i\ell_j) \) relies on five free parameters: 
 \( m_\chi \) and the coefficients \( c_{ik}^v \), \( c_{jk}^v \), \( c_{ik}^a \), \( c_{jk}^a \). Please be aware that when \( c_{ik}^v \) and \( c_{ik}^a \) are equal, the model constraints dictate that \( \Gamma(H\to\ell_i\bar\ell_j) \) vanishes identically. This arises from the fact that if \( c_{ik}^v = c_{ik}^a \), the charges under \( U(1)_\chi \) are generation-independent, precluding any tree-level LFV interaction.
Our focus lies in a scenario where \( \chi \) is a light boson, and the decay \( \mu\to e\chi \) is viable, implying \( m_\chi \leq m_\mu \). Additionally, our attention is directed toward low-energy processes, thus \( \Lambda \) remains below the electroweak scale.

In Eq.~(\ref{eq:BR-Hll}), we illustrate the contribution of a single lepton, $\ell_k$, within the loop. However, it is crucial to note that the interaction $\bar{\ell}_i\ell_j\chi$ can either conserve ($\ell_i=\ell_j$) or violate ($\ell_i\neq\ell_j$) lepton flavor. Consequently, our analysis incorporates three distinct contributions to the triangle diagram: electrons, muons, and taus participating in the loop, along with the associated interferences. However, the interference effects are suppressed and thus neglected. 

\begin{figure}[t!]
\centering
\begin{subfigure}[t]{0.49\textwidth}
\centering
\includegraphics[width=\textwidth]{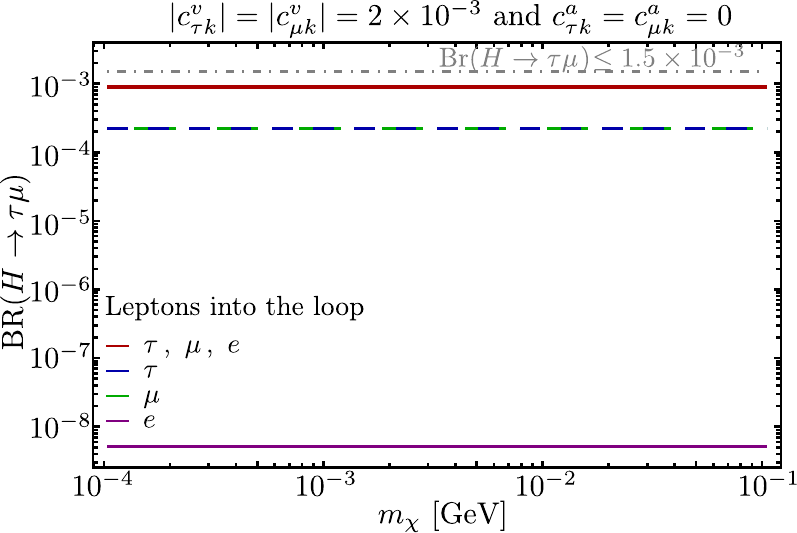}
\caption{${\rm BR}(H\to\tau\mu)$}
\label{fig:Htaumu_c}
\end{subfigure}
\begin{subfigure}[t]{0.49\textwidth}
\centering
\includegraphics[width=\textwidth]{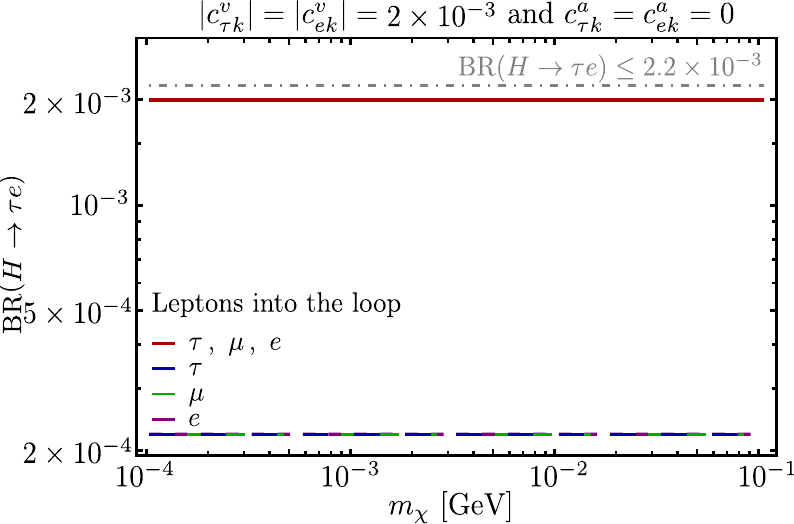}
\caption{${\rm BR}(H\to\tau e)$}
\label{fig:Htaue_c}
\end{subfigure}
\begin{subfigure}[t]{0.49\textwidth}
\centering
\includegraphics[width=\textwidth]{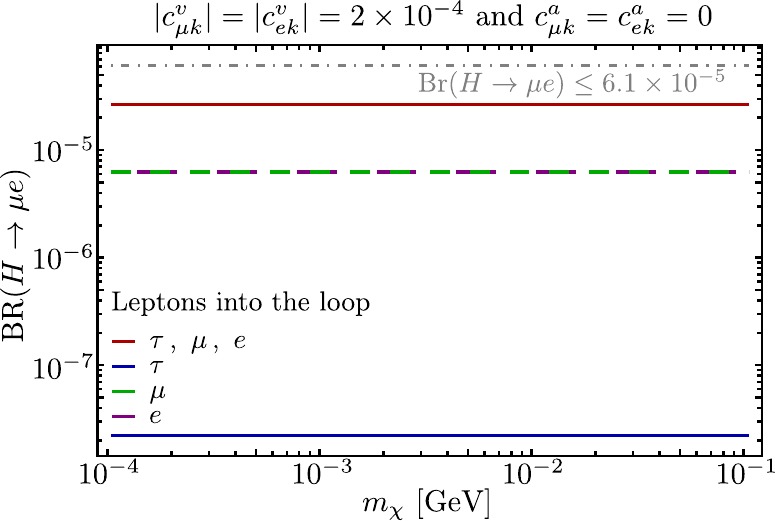}
\caption{${\rm BR}(H\to\mu e)$ }
\label{fig:Hmue_c}
\end{subfigure}
\caption{${\rm BR}(H\to\ell_i \ell_j)$ as a function to $m_\chi~\in~[0,m_\mu]$.}
\label{fig:Hlilj}
\end{figure}

In Figures \ref{fig:Hlilj}, we present the branching ratios ${\rm BR}(H\to\ell_i\ell_j)$ as a function of $m_\chi$ for the three cLFV channels: $H\to\tau\mu$, $H\to\tau e$, and $H\to\mu e$. For our numerical analysis, we consider $m_\chi$ ranging from $0$ to $m_\mu$, and utilize the experimental total Higgs decay width, $\Gamma_H=\left(3.7^{~+1.9}_{~-1.4}\right)$ MeV \cite{Navas:2024prd}. 
To ensure clarity and ease of interpretation, we adopt the simplifying assumptions \( c_{jk}^a = 0 \), thereby suppressing the axial contribution. Suitable values are assigned to \( c_{jk}^v \), as indicated in each figure. The red line corresponds to the scenario where all three lepton contributions are active within the loop, {\it i.e.}, $\tau$, $\mu$, and $e$. Conversely, when only a single lepton contribution is activated into the loop, we represent it with purple, green, and blue lines for $e$, $\mu$, and $\tau$, respectively. To contextualize our findings, the grey line denotes the current upper limit, as tabulated in Table \ref{Table:Br}. Dashed lines are employed to address cases of line overlap, ensuring clarity in the visualization of results.

Our observations demonstrate distinct dominant contributions for each decay process, as depicted in Figures \ref{fig:Htaumu_c}, \ref{fig:Htaue_c}, and \ref{fig:Hmue_c}. For the decay \(H \to \tau \mu\), the diagrams featuring \(\tau\) and \(\mu\) within the loop yield the dominant contributions. In the case of \(H \to \tau e\), all three diagrams, with \(\tau\), \(\mu\), and \(e\) in the loop, exhibit comparable significance. Conversely, for \(H \to \mu e\), the dominant diagrams involve \(\mu\) and \(e\).
This behavior is explained by Eq.~(\ref{eq:BR-Hll}), which states that \(\mathrm{BR}(H \to \ell_i \ell_j) \propto \frac{m_{\ell_k}^2}{m_{\ell_i^k}^2 m_{\ell_j^k}^2}\). The masses in the denominator result from the relationship between the effective couplings \(f_{jk}\) and \(g_{jk}\) with the coefficients \(c_{jk}^v\) and \(c_{jk}^a\) as described in Eq.~(\ref{eq:f-g}). 

If the lepton active is an electron, then \(\mathrm{BR}(H \to \tau \mu) \propto \frac{m_e^2}{m_{\tau}^2 m_{\mu}^2}\). Conversely, if $\mu$ or $\tau$ are active, then \(\mathrm{BR}(H \to \tau \mu) \propto \frac{1}{m_{\tau}^2}\). Given that \(\frac{1}{m_{\tau}^2} \gg \frac{m_e^2}{m_{\tau}^2 m_{\mu}^2}\), it follows that the dominant contribution to this decay channel arises when \(\tau\) and \(\mu\) are active in the loop. 
Similarly, the dominant contribution to the decay \(H \to \mu e\) arises from scenarios where the muon and electron are active in the loop. With these leptons in the loop, \(\mathrm{BR}(H \to \mu e) \propto \frac{1}{m_{\mu}^2}\). Conversely, if the tau lepton is active, \(\mathrm{BR}(H \to \mu e) \propto \frac{1}{m_{\tau}^2}\). Given that \(\frac{1}{m_{\mu}^2} > \frac{1}{m_{\tau}^2}\), it follows that the contribution from the muon and electron is more significant. Lastly, for the decay \(H \to \tau e\), \(\mathrm{BR}(H \to \tau e) \propto \frac{1}{m_{\tau}^2}\) for all possible values of \(m_{\ell_k}\).

Furthermore, it is noteworthy that in Figures \ref{fig:Hlilj}, the branching ratio displays minimal sensitivity to the $\chi$-boson mass and remains finite in the massless limit. It is important to acknowledge that without defining \( f_{jk} \propto m_\chi c_{jk}^v \) and \( g_{jk} \propto m_\chi c_{jk}^a \) (see Eq.~(\ref{eq:f-g})), an unphysical divergence emerges as \( m_\chi \) approaches zero. This divergence would render the results unreliable for a light $\chi$-boson mass \cite{Ibarra:2021xyk}.

By employing the upper limits provided for ${\rm BR}(H\to e\mu\,,e\tau\,,\mu\tau)$ and the flavor-conserving processes ${\rm BR}(H\to ee\,,\mu\mu\,,\tau\tau)$, which are detailed in Table \ref{Table:Br}, we can discern constraints on the individual coefficients $|c_{ik}^{v(a)}|$. Under the assumption that all $|c_{ik}^v|\neq0~{\rm and}~|c_{ik}^a|=0$, and fixing the parameter $m_\chi=m_\mu/2$, we derive restrictions for all $|c_{ik}^v|$.

When analyzing the branching fractions, we exclusively consider the dominant contributions within each process,  as illustrated in Figures \ref{fig:Htaumu_c}, \ref{fig:Htaue_c}, and \ref{fig:Hmue_c} for the cLFV decays. Regarding the flavor-conserving channels $H\to\ell_i\bar\ell_i$, which although are not directly analyzed, can be described by Eq.~(\ref{eq:BR-Hll}), the dominant contributions arise when $m_{\ell_k}=m_{\ell_i}$. This deduction is evident, recalling that ${\rm BR}(H\to\ell_i\ell_i)\propto\frac{m_{\ell_k}^2}{m_{\ell_i^k}^2m_{\ell_i^k}^2}$.
The derived constraint regions for all $|c_{ik}^v|$ are as follows:
\begin{align}\label{eq:CR_cij}
    &0<|c_{\mu\mu}^v|\lesssim5.26\times10^{-4}\,,\quad 0<|c_{\tau\tau}^v|\lesssim8.41\times10^{-3}\,,\quad 0<|c_{ee}^v|\lesssim3.96\times10^{-5}\,,\nonumber\\
    &0<|c_{\mu e}^v|\lesssim5.35\times10^{2}\sqrt{\frac{1}{4.41\times10^{12}+1.59\times10^{19}|c_{ee}^v|^2}}\,,\quad 0<|c_{\tau\mu}^v|\lesssim1.33\times10^{-3}\quad \text{and}\nonumber\\
    &  0<|c_{\tau e}^v|\lesssim1.76\times10^{-3}\sqrt{\frac{4.58\times10^{13}-4.4\times 10^{17}|c_{\mu e}^v|^2}{5.5\times10^{13}+7.78\times10^{17}|c_{ee}^v|^2}}\,.
\end{align}

It is important to emphasize that these constraint regions remain valid across different values of \(m_\chi\), as we have shown that \({\rm BR}(H \to \ell_i \ell_j)\) exhibits minimal dependence on \(m_\chi\). Furthermore, under the assumption that \(c_{ik}^a \neq 0\) and \(c_{ik}^v = 0\), and keeping $m_\chi$ consistent, the resultant constraint regions would remain consistent. This limit on the coefficients \(c_{ii}^v\) can, in turn, be translated into constraints on the Yukawa couplings, \(U(1)_\chi\) charges and gauge coupling, and expectation values of the scalar doublets \(\phi_{ij}\) in the tree-level model, ensuring the correct couplings with the Higgs are reproduced. 
Let us finish this section by noting that while previous studies, such as the one in \cite{Ibarra:2021xyk}, suggest that the most stringent constraints on the coefficients \(c_{jk}^{v(a)}\) would likely arise from lepton-flavor violating three-body decays, such as \(\mu \to 3e\) and \(\tau \to 3\ell\) (with \(\ell = e, \mu\)), the focus of this work lies elsewhere. Our aim is not to derive the most restrictive bounds on these coefficients but to explore the unique lepton-flavor violating Higgs decays mediated by an ultralight gauge boson. We seek to investigate the phenomenology of these decays in the Higgs sector, providing observables that are sensitive to the mass of the \(\chi\) boson, as we will discuss in the next section.

\subsection[Higgs decays with $\chi$ on-shell]{Higgs decays with $\boldsymbol{\chi}$ on-shell}\label{On-S}

\begin{figure}[t!]
\centering
\includegraphics[width=0.8\textwidth]{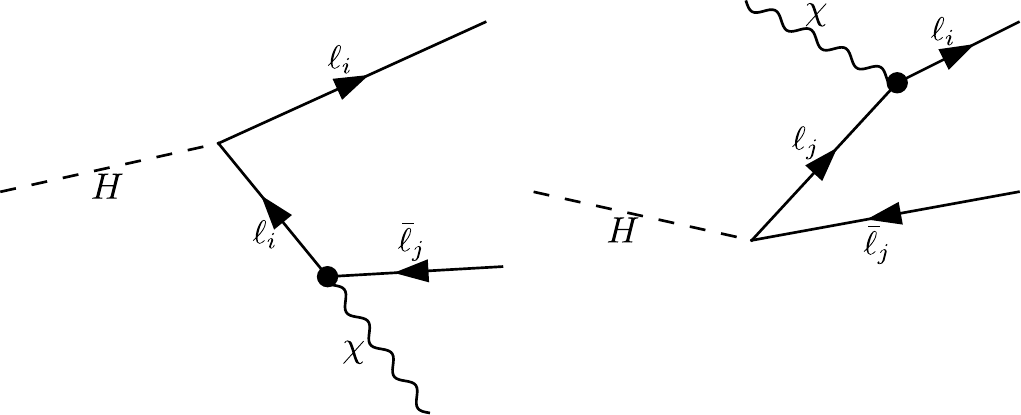} 
\caption{Contribution to $H\to\ell_i\bar{\ell}_j\chi$. Here $\ell_i\,,\ell_j=e\,,\mu\,,\tau$ and $\ell_i=\ell_j$ or $\ell_i\neq\ell_j$.}
\label{fig:hllchi}
\end{figure}

Utilizing the effective Lagrangian outlined in Eq.~(\ref{eq:Eff-Lag}), we can induce the decays $H\to\ell_i\bar{\ell}_j\chi$ at the tree level, as depicted in Fig.~\ref{fig:hllchi}. Here, we introduce the Mandelstam variables $t\equiv(q_{\ell_j}+q_{\chi})^2$ and $s\equiv(q_{\ell_i}+q_{\chi})^2$, where $q_{\ell_j}$, $q_{\ell_i}$, and $q_{\chi}$ denote the four-momenta of $\ell_j$, $\ell_i$, and $\chi$, respectively. The differential decay rate is then expressed as:
\begin{equation}\label{eq:dG-Hllchi}
    \frac{d^2\Gamma(H\to\ell_i\bar{\ell}_j\chi)}{ds~ dt}=\frac{1}{32(2\pi)^3 M_H^3}\overline{|\mathcal{M}_{H\to\ell_i\bar{\ell}_j\chi}(s,t)|^2}\,,
\end{equation}
where the squared amplitude as a function of $s$ and $t$ is given in Appendix \ref{apx:SA-chi-off}.
The kinematic limits are
\begin{align}
    t_{\binom{{\rm max}}{{\rm min}}}=\frac{1}{4s}\biggl[\left(M_H^2+m_\chi^2-m_{\ell_i}^2-m_{\ell_j}^2\right)^2-\left(\lambda^{1/2}[s,m_{\ell_i}^2,m_\chi^2]\mp\lambda^{1/2}[s,M_H^2,m_{\ell_j}^2]\right)^2\biggr]\,,
\end{align}
and $(m_{\ell_i}+m_\chi)^2\leq s \leq (M_H-m_{\ell_j})^2$, where $\lambda[s,M_H^2,m_{\ell_j}^2]$ is the usual Käll\'en function. 

By leveraging the constraint regions on the coefficients $|c_{ik}^v|$ outlined in Eqs.~(\ref{eq:CR_cij}), we can establish indirect upper bounds on ${\rm BR}(H\to\ell_i\ell_j\chi)$, which is defined as:
\begin{equation}
    {\rm BR}(H\to\ell_i\ell_j\chi)=\frac{\Gamma(H\to\ell_i\bar{\ell}_j\chi)+\Gamma(H\to\bar{\ell}_i\ell_j\chi)}{\Gamma_H}\,.
\end{equation}

\begin{figure}[t!]
\centering
\includegraphics[width=0.70\textwidth]{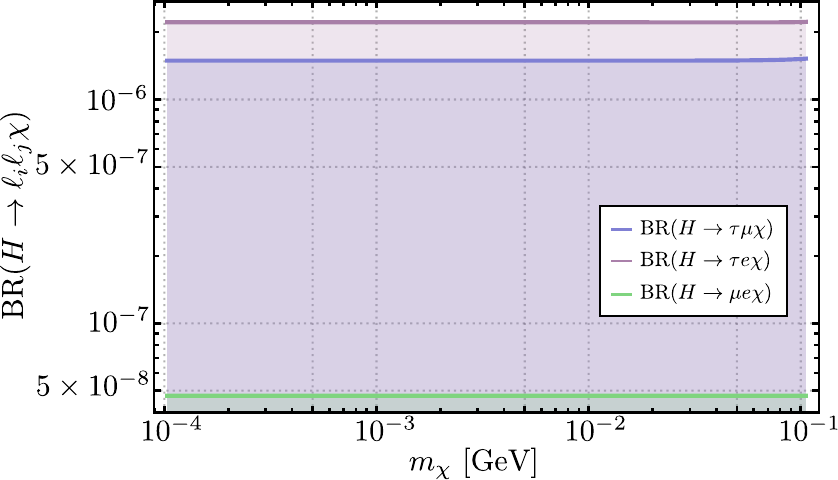}
\caption{Upper bound on ${\rm BR}(H\to\ell_i\ell_j\chi)$ as a function to  $m_\chi \in [0,m_\mu]$. We use the constraint regions in Eqs.~(\ref{eq:CR_cij}) for the $|c_{ij}^v|$.}
\label{fig:BR_HLlchi}
\end{figure}

In Figure \ref{fig:BR_HLlchi}, we present the upper bound on ${\rm BR}(H\to\ell_i\ell_j\chi)$ across the range $m_\chi \in [0,m_\mu]$. These bounds are derived from constraints established by the upper limits of $H\to\ell_i\bar\ell_j$ decays, as outlined in Eqs.~(\ref{eq:CR_cij}), while assuming $c_{ik}^a=0$. Notably, similar to the ${\rm BR}(H\to\ell_i\ell_j)$ decays, the 3-body Higgs decay ${\rm BR}(H\to\ell_i\ell_j\chi)$ displays minimal dependence on the $\chi$-boson mass. 
The upper limits on \({\rm BR}(H\to\ell_i\ell_j\chi)\) are several orders of magnitude more stringent than those reported in experimental searches for Higgs decays \cite{Navas:2024prd}, particularly for the decay \(H\to\mu e\chi\). Detecting such decays would require substantial experimental effort, including effectively managing the decay background allowed by the Standard Model, such as \(H\to\ell_i\bar{\ell}_i\gamma\), especially when the photon has very low energy.

\subsubsection{Lepton Energy Spectrum}
Analyzing the lepton energy spectrum for these processes, $H\to\ell_i\bar{\ell}_j\chi$ is also interesting. In the Higgs rest frame, $\vec{p}_{H}=\vec{0}$ then $E_{H}=M_H$, where $\vec{p}_H$ and $E_H$ represent the momentum and energy of the Higgs, respectively. Consequently, the Mandelstam variable $t=M_{H}^2+m_{\ell_i}^2-2M_{H}E_{\ell_i}$. By applying a variable change to Eq.~(\ref{eq:dG-Hllchi}), we can express the partial decay rate as a function of $s$ and $E_{\ell_i}$ as follows:
\begin{equation}\label{eq:dG-Hllchi-sEi}
    \frac{d^2\Gamma(H\to\ell_i\bar{\ell}_j\chi)}{dsdE_{\ell_i}}=\frac{1}{(2\pi)^3 16M_H^2}\overline{|\mathcal{M}_{H\to\ell_i\bar{\ell}_j\chi}(s,E_{\ell_i})|^2}\,,
\end{equation}
where
\begin{align*}
    m_{\ell_i}\leq E_{\ell_i}\leq \frac{M_H^2+m_{\ell_i}^2-(m_{\ell_j}+m_\chi)^2}{2M_H}.
\end{align*}

\begin{figure}[t!]
\centering
\begin{subfigure}[t]{0.49\textwidth}
\centering
\includegraphics[width=\textwidth]{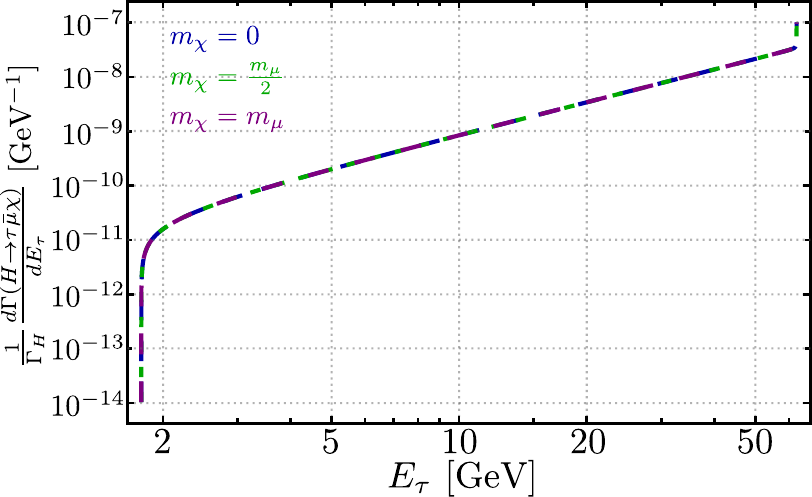}
\caption{$H\to\tau\bar{\mu}\chi$ as function of $E_\tau$.}
\label{fig:ES_Hmuechi}
\end{subfigure}
\begin{subfigure}[t]{0.49\textwidth}
    \centering
    \includegraphics[width=\textwidth]{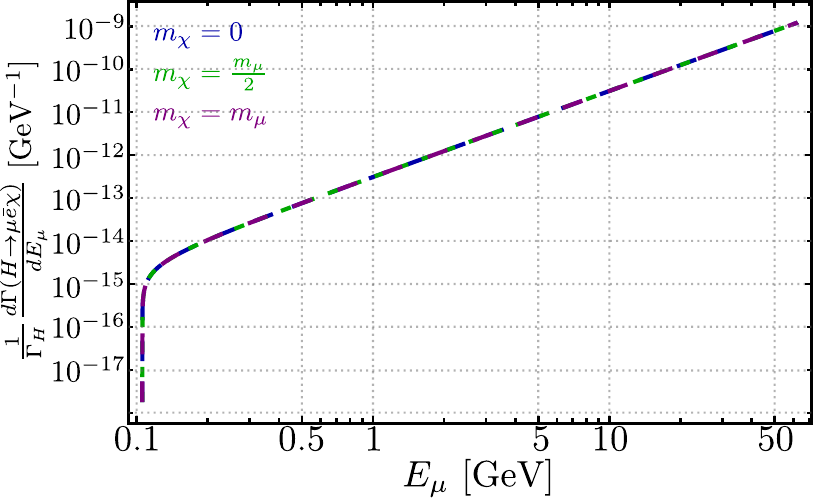}
    \caption{$H\to\mu \bar{e}\chi$ as function of $E_\mu$.}
    \label{fig:ES-Hmuechi}
\end{subfigure}
\caption{Energy spectrum $\frac{1}{\Gamma_H}\frac{d\Gamma(H\to\ell_i\bar{\ell}_j\chi)}{dE_{\ell_i}}$ as a function of $E_{\ell_i}$. We assume $c_{ij}^a=0$ while $|c_{ij}^v|$ follows the constraints derived in Eqs.~(\ref{eq:CR_cij}). We consider three options for the $\chi$-boson mass: $m_\chi=0\,,$ $m_\chi=m_\mu/2$, and $m_\chi=m_\mu$.}
\label{fig:Energy-Spectrum}
\end{figure}

In Figure \ref{fig:Energy-Spectrum}, we present the lepton energy spectrum $\frac{1}{\Gamma_H}\frac{d\Gamma(H\to\ell_i\bar{\ell}_j\chi)}{dE_{\ell_i}}$ as a function of $E_{\ell_i}$, where \(\Gamma_H\) denotes the experimental total Higgs decay width.  This figure illustrates the decays  $H\to\tau\bar{\mu}\chi$ and $H\to\mu \bar e\chi$. The lepton energy spectrum for $H\to\tau \bar e\chi$ resembles that shown in Fig.~\ref{fig:ES_Hmuechi} for the analogous decay $H\to\tau\bar\mu\chi$. We set \(c_{ij}^a=0\) and  apply the upper constraints  for \(|c_{ij}^v|\) as outlined in Eqs.~(\ref{eq:CR_cij}). We explore three different choices of $\chi$-boson mass: $m_\chi=0$ (blue line), $m_\chi=m_\mu/2$ (green line), and $m_\chi=m_\mu$ (purple line). Notably, all scenarios exhibit overlapping spectra, indicating that the decays $H\to\ell_i\bar{\ell}_j\chi$ display minimal dependence on the $\chi$-boson mass. This outcome is consistent with the observations in Fig.~\ref{fig:BR_HLlchi} and the previous analysis of processes with $\chi$ off-shell: $H\to\ell_i\bar\ell_j$, as illustrated in Figs.~\ref{fig:Hlilj}.

\subsubsection{Angular Observables}
We also examined the decays \(H\to\ell_i\bar\ell_j\chi\) as functions of lepton energy (\(E_{\ell_i}\)) and an angular variable \(\cos{\theta_{\ell_i\ell_j}}\). Here, \(\theta_{\ell_i\ell_j}\) represents the angle between the momenta of the two leptons in the rest frame of the \(\ell_i-\chi\) system, where \(\vec{q}_{\ell_i}+\vec{q}_\chi=\vec{0}\). Consequently, we have 
\(|\vec{p}_H|=|\vec{q}_{\ell_j}|=\sqrt{E_H^2-M_H^2}\) and \(|\vec{q}_{\ell_i}|=|\vec{q}_\chi|=\sqrt{E_{\ell_i}^2-m_{\ell_i}^2}\), with
\[
E_H=\frac{(E_{\ell_i}+E_\chi)^2+M_H^2-m_{\ell_j}^2}{2 (E_{\ell_i}+E_\chi)}\,.
\]

Then \(s=(E_{\ell_i}+E_\chi)^2\) and 
\(t=m_{\ell_j}^2+m_\chi^2+2\left(E_{\ell_j}E_\chi+|\vec{q}_{\ell_i}||\vec{q}_{\ell_j}|\cos{\theta_{\ell_i\ell_j}}\right)\). Thus, the partial decay rate as a function of \(E_{\ell_i}\) and \(\cos{\theta_{\ell_i\ell_j}}\) can be expressed as:
\begin{align}
\frac{d^2\Gamma(H\to\ell_i\bar{\ell}_j\chi)}{dE_{\ell_i} d\cos{\theta_{\ell_i\ell_j}}}=&\frac{(E_\chi+E_{\ell_i})^2|\vec{q}_{\ell_i}||\vec{q}_{\ell_j}|}{(2\pi)^3~ 8M_H^3 E_\chi}\overline{|\mathcal{M}_{H\to\ell_i\bar{\ell}_j\chi}(\cos{\theta_{\ell_i\ell_j}},E_{\ell_i})|^2}\,.
\end{align}

We explored various angular observables for the processes with $\chi$ on-shell: \(H\to\tau\mu\chi\,,\) \(H\to\tau e\chi\,,\) \(H\to\mu e\chi\). For the following analyses, we assumed $c_{ij}^a=0$ and determined the values of $|c_{ij}^v|$ based on the constraints presented in Eqs.~(\ref{eq:CR_cij}). Additionally, we considered three scenarios for the \(\chi\)-boson mass: \( m_\chi = 0 \), \( m_\chi = m_\mu/2 \), and \( m_\chi \sim m_\mu \). 

\begin{figure}[t!]
    \centering
    \begin{subfigure}[b]{0.3\textwidth}
     \centering
     \includegraphics[width=\textwidth]{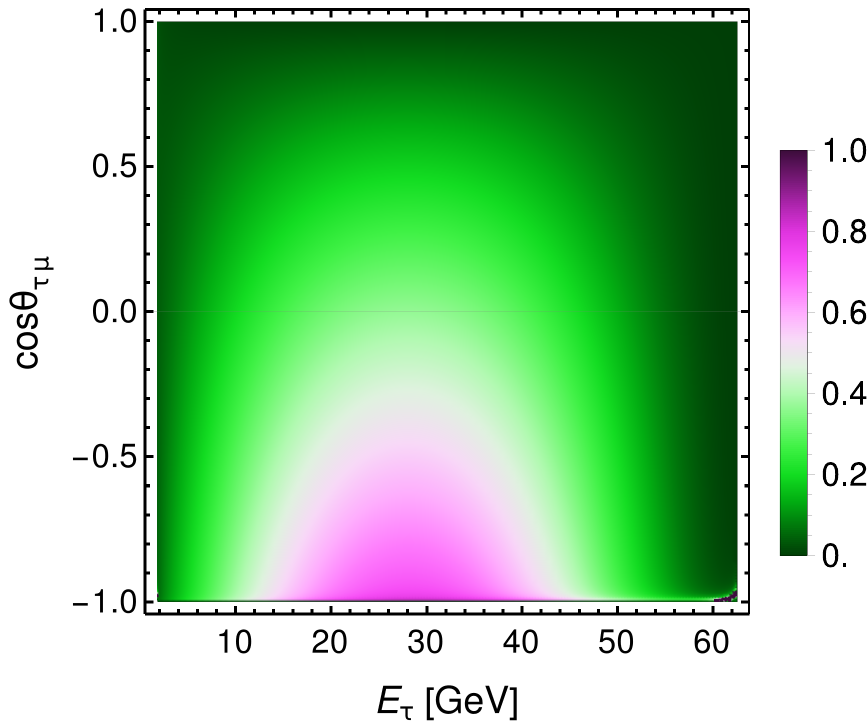}
     \caption{$m_\chi=0$}
    \end{subfigure}
    \hfill
    \begin{subfigure}[b]{0.3\textwidth}
     \centering
     \includegraphics[width=\textwidth]{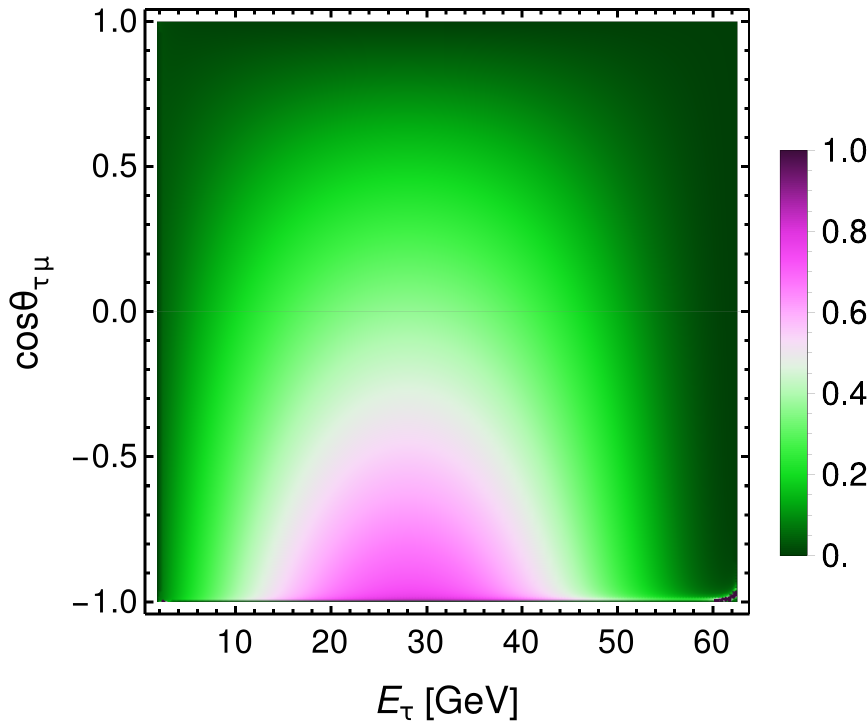}
     \caption{$m_\chi=\frac{m_\mu}{2}$}
    \end{subfigure}
    \hfill
    \begin{subfigure}[b]{0.3\textwidth}
     \centering
     \includegraphics[width=\textwidth]{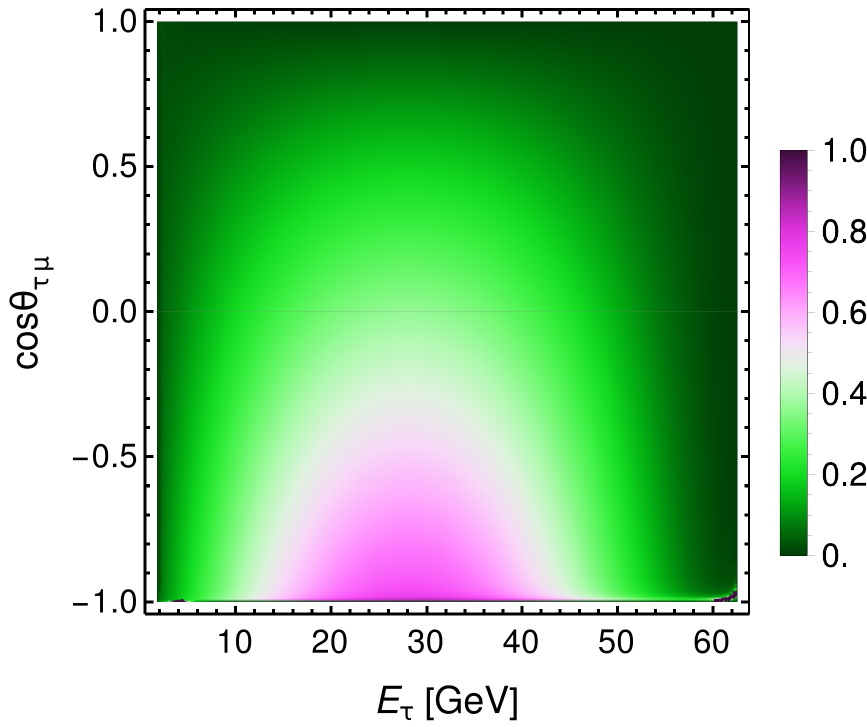}
     \caption{$m_\chi\sim m_\mu$}
    \end{subfigure}
    \caption{Dalitz Plot of the decay $H\to\tau\bar{\mu}\chi$ as a function of $E_{\tau}$ and $\cos\theta_{\tau\mu}$.}
    \label{fig:DP-Htaumuchi}
\end{figure}

In Figure \ref{fig:DP-Htaumuchi}, we illustrate the normalized partial decay rate \( \frac{d^2\Gamma(H\to\ell_i\bar{\ell}_j\chi)}{dE_{\ell_i} d\cos{\theta_{\ell_i\ell_j}}} \) as a Dalitz plot. The distributions of Dalitz plots for the other two processes, $H\to\tau\bar{e}\chi$ and $H\to\mu\bar{e}\chi$, resemble that shown in Fig.~\ref{fig:DP-Htaumuchi}. Please be aware that the Dalitz plot distribution is not sensitive to the \(\chi\)-boson mass, consistent with the other observables explored.

To investigate whether differential distributions in \(H\to\ell_i\bar\ell_j\chi\) decays can provide insights into the mass of the \(\chi\)-boson, we consider the Lepton Charge Asymmetry as a function of \(\cos\theta_{\ell_i\ell_j}\). This asymmetry is defined as follows:
\begin{equation}\label{Eq:LCA}
    \mathcal{A}^{L-C}(H\to\ell_i\ell_j\chi) = \frac{\frac{d\Gamma(H\to\ell_i\bar{\ell}_j\chi)}{d\cos\theta_{\ell_i \ell_j}} - \frac{d\Gamma(H\to\bar{\ell}_i\ell_j\chi)}{d\cos\theta_{\ell_i \ell_j}}}{\frac{d\Gamma(H\to\ell_i\bar{\ell}_j\chi)}{d\cos\theta_{\ell_i \ell_j}} + \frac{d\Gamma(H\to\bar{\ell}_i\ell_j\chi)}{d\cos\theta_{\ell_i \ell_j}}}\,.
\end{equation}
\begin{figure}[t!]
\centering
\begin{subfigure}[t]{0.49\textwidth}
\centering
\includegraphics[width=\textwidth]{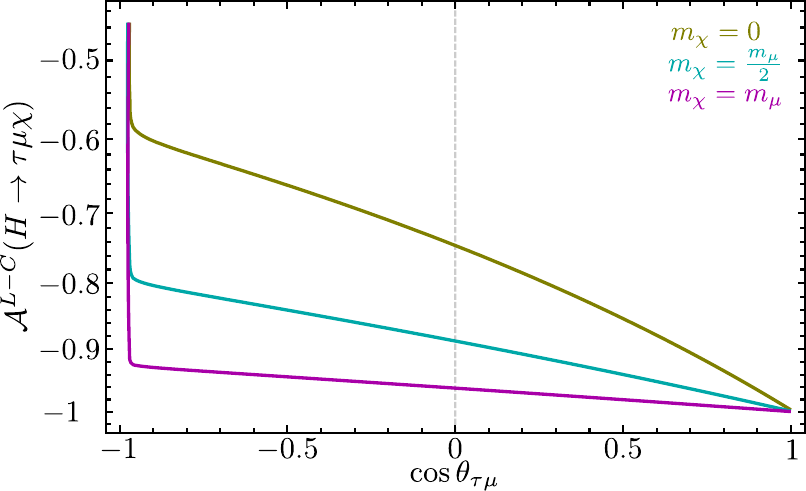}
\caption{$H\to\tau\mu\chi$ as function of $\cos\theta_{\tau\mu}$.}
\label{fig:LCA_taumu}
\end{subfigure}
\begin{subfigure}[t]{0.49\textwidth}
    \centering
    \includegraphics[width=\textwidth]{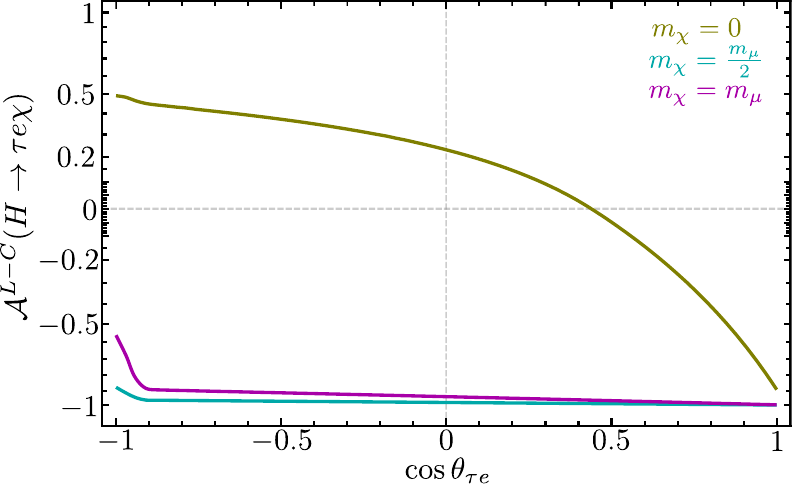}
    \caption{$H\to\tau e\chi$ as function of $\cos\theta_{\tau e}$.}
    \label{fig:LCA_taue}
\end{subfigure}
\begin{subfigure}[b]{0.49\textwidth}
    \centering
    \includegraphics[width=\textwidth]{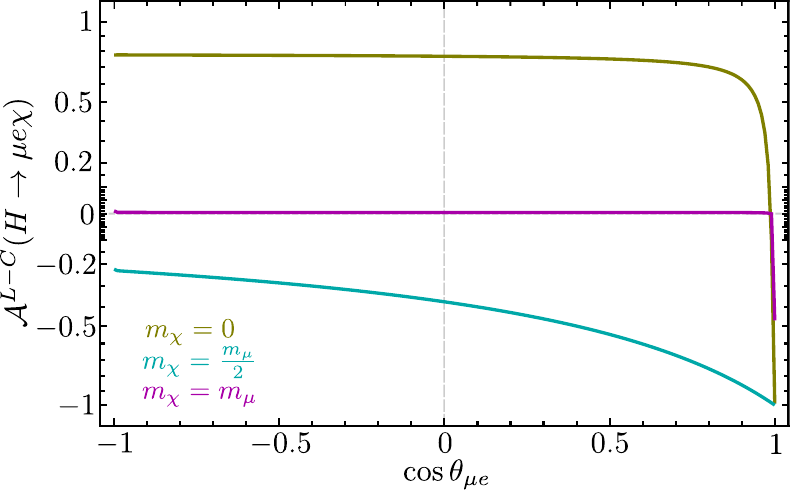}
    \caption{$H\to\mu e\chi$ as function of $\cos\theta_{\mu e}$.}
    \label{fig:LCA_mue}
\end{subfigure}
\caption{Lepton Charged Asymmetry $\mathcal{A}^{L-C}(H\to\ell_i\ell_j\chi)$ as a function of $\cos\theta_{\ell_i \ell_j}$. We assume $c_{ij}^a=0$ while $|c_{ij}^v|$ follows the constraints derived in Eqs.~(\ref{eq:CR_cij}). We consider three options for the $\chi$-boson mass:  \(m_\chi=0\) (olive line), \(m_\chi=m_\mu/2\) (cyan line), and \(m_\chi=m_\mu\) (magenta line).}
\label{fig:LCA}
\end{figure}

This observable is plotted in Fig.~\ref{fig:LCA} for the lepton flavor-violating channels: \(H \to \tau \mu \chi\), \(H \to \tau e \chi\), \(H \to \mu e \chi\). In Fig.~\ref{fig:LCA}, we took the heaviest lepton as the particle (\(\ell_i\)) in the positive partial decay rate for the asymmetry calculations. To maintain consistency with the definition in Eq.~(\ref{Eq:LCA}), we list the heaviest lepton first in the notation.

For decays involving the lepton \(\tau\), the asymmetry \(\mathcal{A}^{L-C}\) is generally negative, except when \(m_\chi \sim 0\) in the decay \(H\to\tau e\chi\). The asymmetry \(\mathcal{A}^{L-C}\) is close to zero for \(H\to\mu e \chi\) when \(m_\chi\) is large but becomes negative at \(m_\chi = 0\) and positive for intermediate values of the \(\chi\)-boson mass. As shown in Fig.~\ref{fig:LCA_taumu}, for \(H\to\tau\mu\chi\), the asymmetry can distinguish between intermediate and large values of \(m_\chi\) with about $9\%$ precision, and between small and large values with about $25\%$ precision for small values of \(\cos\theta_{\tau\mu}\). In the decay \(H\to\tau e\chi\), this observable is quite sensitive to the \(\chi\)-boson mass for small values of \(m_\chi\) but can differentiate between intermediate and large values with about $5\%$ precision. 
The decay \(H\to\mu e\chi\) shows the greatest sensitivity to \(m_\chi\), particularly for large values of \(\cos\theta_{\mu e}\). Remarkably, the Lepton Charge Asymmetry exhibits significant sensitivity to the \(\chi\)-boson mass, especially when \(m_\chi \sim 0\) and in decays involving an electron, such as \(H\to\tau e\chi\) and \(H\to\mu e\chi\). 

We also analyzed the Forward-Backward Asymmetry as a function to $E_{\ell_i}$ defined as follow
\begin{equation}
    \mathcal{A}^{F-B}(H\to\ell_i\bar\ell_j \chi)=\frac{\int_{-1}^0 \frac{d\Gamma(H\to\ell_i\bar\ell_j \chi)}{dE_{\ell_i}d\cos\theta_{\ell_i\ell_j}}d\cos\theta_{\ell_i\ell_j}-\int_{0}^1 \frac{d\Gamma(H\to\ell_i\bar\ell_j \chi)}{dE_{\ell_i}d\cos\theta_{\ell_i\ell_j}}d\cos\theta_{\ell_i\ell_j}}{\int_{-1}^0 \frac{d\Gamma(H\to\ell_i\bar\ell_j \chi)}{dE_{\ell_i}d\cos\theta_{\ell_i\ell_j}}d\cos\theta_{\ell_i\ell_j}+\int_{0}^1 \frac{d\Gamma(H\to\ell_i\bar\ell_j \chi)}{dE_{\ell_i}d\cos\theta_{\ell_i\ell_j}}d\cos\theta_{\ell_i\ell_j}}
\end{equation}

\begin{figure}[t!]
\centering
\begin{subfigure}[t]{0.49\textwidth}
\centering
\includegraphics[width=\textwidth]{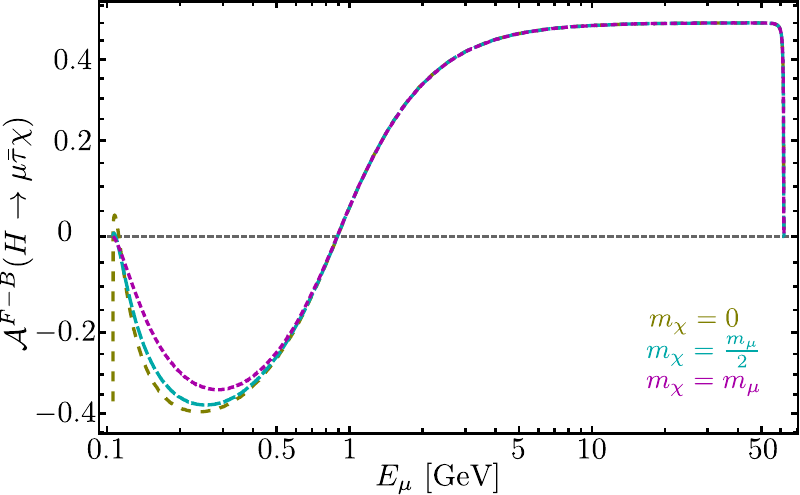}
\caption{$H\to\mu\bar\tau\chi$ as function of $E_{\mu}$.}
\label{fig:FBA_taumu}
\end{subfigure}
\begin{subfigure}[t]{0.49\textwidth}
    \centering
    \includegraphics[width=\textwidth]{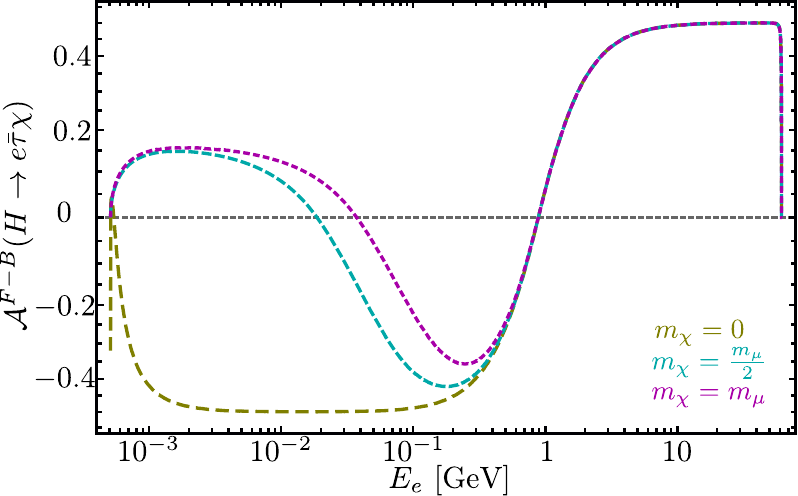}
    \caption{$H\to e\bar\tau\chi$ as function of $E_e$.}
    \label{fig:FBA_taue}
\end{subfigure}
\begin{subfigure}[b]{0.49\textwidth}
    \centering
    \includegraphics[width=\textwidth]{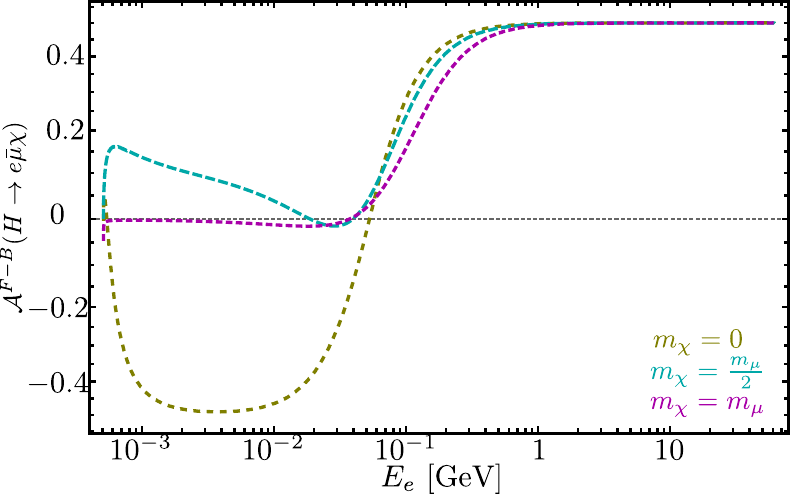}
    \caption{$H\to e\bar\mu\chi$ as function of $E_e$.}
    \label{fig:FBA_mue}
\end{subfigure}
\caption{Forward-Backward Asymmetry $\mathcal{A}^{F-B}(H\to\ell_i\bar\ell_j\chi)$ as a function of $E_i$. We assume $c_{ij}^a=0$ while $|c_{ij}^v|$ follows the constraints derived in Eqs.~(\ref{eq:CR_cij}). We consider three options for the $\chi$-boson mass:\(m_\chi=0\) (olive line), \(m_\chi=m_\mu/2\) (cyan line), and \(m_\chi=m_\mu\) (magenta line).}
\label{fig:FBA}
\end{figure}

We illustrate this observable in Fig.~\ref{fig:FBA} as a function of \(E_{\ell_i}\). It is apparent that this observable is sensitive to the \(\chi\)-boson mass, particularly in the low energy range. This sensitivity is more pronounced in channels involving an electron and when \(\chi\) is ultralight. For the channel \(H\to\bar{\mu}\tau\chi\), the sensitivity is slight; at best, there is a \(\sim 10\%\) difference between the lightest and heaviest values of the \(\chi\)-boson. In the decays \(H \to e\bar\tau \chi\) and \(H \to e\bar\mu \chi\), these observables exhibit significant sensitivity to the \(\chi\)-boson mass at small \(m_\chi\) values. Additionally, in the best-case scenario, they can distinguish between intermediate and large \(m_\chi\) values with a precision of about $15\%$ and $13\%$, respectively.

Remarkably, the asymmetries are the only analyzed observables that exhibit sensitivity to the \(\chi\)-boson mass. While this analysis assumes a background-free scenario, it is important to note that achieving the required precision for such measurements, particularly for \(H\to\ell_i\bar\ell_j\gamma\) decays, would be challenging. Attaining this precision would necessitate meticulous and dedicated analysis to ensure accurate results.

\section{Conclusions}\label{Conclusions}

Lepton Flavor Violation is highly suppressed in the SM extended with right-handed neutrinos, but it may manifest at measurable rates in several well-motivated BSM models. The discovery of the Higgs boson at the LHC has introduced a new avenue for exploring LFV in the decays of this scalar boson.

In this study, we explored the influence of an ultralight gauge boson, \(\chi\), in mediating cLFV transitions. Our analysis incorporates a model that generates the lepton flavor-violating interaction \(\bar{\ell}_i\ell_j\chi\) at the tree level into an effective field theory, ensuring the \(\chi\)-boson mass remains well-defined as it approaches zero. We examined the LFV Higgs decay phenomenology under both on-shell and off-shell \(\chi\) conditions. The results indicate that these processes exhibit minimal dependence on the \(\chi\)-boson mass, due to the established conditions for the observables to remain finite in the massless limit.

However, the Lepton Charge and Forward-Backward Asymmetries, display sensitivity to \(m_\chi\). These asymmetries indirectly reveal information about the range of the \(\chi\)-boson mass as a function of the lepton energy or the angle \(\theta_{\ell_i \ell_j}\). This sensitivity provides a potential method for constraining the mass of the \(\chi\)-boson in experimental observations.

Utilizing the upper bounds and measurements of Higgs decays into leptons, \(H \to \ell_i \bar\ell_j\) from CMS and ATLAS Collaborations, we derived indirect upper limits on the decays \(H \to \ell_i \bar\ell_j \chi\). These constraints provide critical insights into the potential signatures of LFV in Higgs decays mediated by an ultralight gauge boson. Collider experiments with enhanced sensitivity to LFV Higgs decay could potentially uncover the presence of an ultralight gauge boson. 

\section*{Acknowledgments}
The research presented herein has been supported by the UNAM Postdoctoral Program (POSDOC) and the PAPIIT project IN102122. We are grateful to Pablo Roig for his invaluable comments and feedback.
\appendix
\section{Appendix: mixing angles and masses in the tree level model}\label{apx:masses}
In this appendix, we provide expressions for the $\chi$-boson mass, the masses of the leptons ($m_\mu$ and $m_e$), and the mixing angles ($\sin2\theta_L$ and $\sin2\theta_R$) as functions of $y_{jk}$, $v_{jk}$, and $q_{\phi_{jk}}$ within the context of the tree-level model.
After the spontaneous symmetry breaking of the symmetry $U(1)_\chi$, the non-zero expectation values for $\phi_{jk}$ generate a mass for the $\chi$ boson: 
\begin{align}
m^2_\chi=g_\chi^2 (q_{\phi_{11}}^2 v_{11}^2+q_{\phi_{12}}^2 v_{12}^2+q_{\phi_{21}}^2 v_{21}^2+q_{\phi_{22}}^2 v_{22}^2)\;.
\label{eq:chi-mass}
\end{align}
The expectation value of the doublet scalars generates a mass term for the charged leptons, $-{\cal L}_{\rm mass}\supset \overline{e_{L_j}} M_{jk} e_{R_k}+{\rm h.c.}$, with 
\begin{align}
M=\begin{pmatrix}
y_{11} v_{11} & y_{12} v_{12} \\
y_{21} v_{21} & y_{22} v_{22} \\
 \end{pmatrix}\;.
 \label{eq:mass_matrix}
\end{align}
We now rotate the fields to express the Lagrangian on the mass eigenstate basis:
\begin{align}
\begin{pmatrix} e_L \\ \mu_L \end{pmatrix}= \begin{pmatrix} \cos\theta_L & \sin\theta_L \\-\sin\theta_L & \cos\theta_L \end{pmatrix} \begin{pmatrix} e_{L_1} \\ e_{L_2} \end{pmatrix}\,,~~~~~~~
\begin{pmatrix} e_R \\ \mu_R \end{pmatrix}= \begin{pmatrix} \cos\theta_R & \sin\theta_R \\-\sin\theta_R & \cos\theta_R \end{pmatrix} \begin{pmatrix} e_{R_1} \\ e_{R_2} \end{pmatrix}
\end{align}
so that $-{\cal L}_{mass}\supset \overline{e_{L}} m_e e_R +\overline{\mu_{L}} m_\mu \mu_R\,+$ h.c., with 
\begin{align}
m_\mu^2&\simeq y_{11}^2v_{11}^2+y_{12}^2v_{12}^2 +y_{21}^2v_{21}^2 +y_{22}^2v_{22}^2 \;,\nonumber \\
m_e^2&\simeq \frac{(y_{11} v_{11} y_{22} v_{22}-y_{12} v_{12} y_{21} v_{21})^2}{y_{11}^2v_{11}^2+y_{12}^2v_{12}^2 +y_{21}^2v_{21}^2 +y_{22}^2v_{22}^2}\;, \nonumber\\
\sin2\theta_L&\simeq-2\frac{y_{11} v_{11} y_{21} v_{21}+y_{12} v_{12} y_{22} v_{22}}{y_{11}^2v_{11}^2+y_{12}^2v_{12}^2 +y_{21}^2v_{21}^2 +y_{22}^2v_{22}^2}\;,\nonumber\\
\sin2\theta_R&\simeq -2\frac{y_{11} v_{11} y_{12} v_{12}+y_{21} v_{21} y_{22} v_{22}}{y_{11}^2v_{11}^2+y_{12}^2v_{12}^2 +y_{21}^2v_{21}^2 +y_{22}^2v_{22}^2}\;,
\label{eq:eigensystem}
\end{align}
where we have used that empirically $m_\mu\gg m_e$.

\section{Appendix: Counterterms for the Amplitude of $H\to\ell_i\bar{\ell}_j$}\label{apx:UV-Divergent}
The amplitude generated by the triangle diagram for the decay \(H \to \ell_i \bar{\ell}_j\) includes UV-divergent terms, necessitating the introduction of counterterms to absorb these divergences. The one-loop amplitudes and the corresponding form factors are initially expressed in terms of Passarino-Veltman functions (see Eq. (\ref{Eq:Loop-Function})). For numerical calculations, however, we work with the renormalized amplitude, where the divergences are canceled through counterterms. The associated counterterm Lagrangian is given by
\begin{align}\label{Eq:LCT}
    \mathcal{L}_{{\rm CT}} = C_{f_{ij}}\bar{\ell}_i\ell_j H + C_{g_{ij}}\bar{\ell}_i\gamma_5\ell_j H + {\rm h.c.}\,,
\end{align}
where the coefficients of the scalar and pseudoscalar operators are specified as:
\begin{align}
  C_{f_{ij}} &= \frac{m_{\ell_k}}{2 m_\chi^2 v \bar{\epsilon}_{{\rm uv}}}\left(M_H^2 - 6m_{\ell_k}^2 + 6 m_\chi^2\right) (f_{ik} f_{jk} - g_{ik} g_{jk})\,, \nonumber\\
  C_{g_{ij}} &= \frac{m_{\ell_k}}{2 m_\chi^2 v \bar{\epsilon}_{{\rm uv}}}\left(M_H^2 - 6 m_{\ell_k}^2 + 6 m_\chi^2\right) (f_{ik} g_{jk} - f_{jk} g_{ik})\,,
\end{align}
with the effective couplings \(f_{ij}\) and \(g_{ij}\) defined in Eq. (\ref{eq:f-g}), and $\frac{1}{\bar{\epsilon}_{{\rm uv}}}\equiv \frac{1}{\epsilon_{{\rm uv}}}-\gamma_E+\ln 4\pi$. The amplitude is renormalized using the \(\overline{\rm MS}\)-scheme, ensuring that only finite contributions remain in the final calculation.

It is important to observe that the counterterm Lagrangian in Eq. (\ref{Eq:LCT}) comprises two distinct contributions: the first term corresponds to a scalar operator, while the second term represents a pseudoscalar operator. The pseudoscalar term does not contribute to the amplitude in the flavor-conserving scenario. This indicates that quantum corrections primarily generate the scalar operator, similar to the behavior observed in the Standard Model.

\section{Appendix: Squared Amplitude}\label{apx:SA-chi-off}

In this appendix, we provide the squared amplitude for the decay with $\chi$ on-shell, \( H\to\ell_i \bar\ell_j\chi \), as a function of the Mandelstam variables \( s \) and \( t \):

{\small 
\begin{align}\label{eq:M2-Hllchi}
\overline{|\mathcal{M}_{H\to\ell_i\bar{\ell}_j\chi}(s,t)|^2}&\simeq  \frac{2}{m_{\ell_i^j}^2 v^2} \Biggl [\frac{m_{\ell_j}^2}{\Gamma_{\ell_j}^2 m_{\ell_j}^2 + (m_{\ell_j}^2 - s)^2} 
\biggl[ \left( 
\left| c^a_{ij} \right|^2 + \left| c^v_{ij} \right|^2 \right) 
\Bigl( 
M_H^2 (m_{\ell_i}^4 + m_{\ell_i}^2 (m_\chi^2 - 2 s) \nonumber\\
&+ m_{\ell_j}^2 m_\chi^2 - 2 m_\chi^4 + s^2) - m_{\ell_i}^4 (3 m_{\ell_j}^2 + s) + 
m_{\ell_i}^2 (m_{\ell_j}^4 - m_{\ell_j}^2 (2 m_\chi^2 - 5 s + t) \nonumber\\
&+s (-2 m_\chi^2 + 2 s + t)) + t (m_{\ell_j}^2 - s) (s - 2 m_\chi^2) + (m_\chi^2 - s) ((m_{\ell_j}^2 + s)^2 + 8 m_{\ell_j}^2 m_\chi^2) \Bigr) \nonumber\\
&+ 6m_{\ell_i} m_{\ell_j} m_\chi^2 \left( \left| c^a_{ij} \right|^2 - \left| c^v_{ij} \right|^2\right) ( M_h^2  - 2 (m_{\ell_j}^2 + s))\biggr]+
{\scriptstyle \left \{  \begin{matrix}  s\leftrightarrow t\\ m_{\ell_i}\leftrightarrow m_{\ell_j}\\  \Gamma_{\ell_j}\rightarrow\Gamma_{\ell_i}
\end{matrix} \right \} }
\Biggr ]
\end{align}}
\normalsize

In this expression, the interference terms are subdominant and have been neglected. Here, \(\Gamma_{\ell_k}\) denotes the total decay width of the lepton \(\ell_k\). It is important to note that the squared amplitude does not exhibit divergences for \(m_\chi\), ensuring the finiteness of the decay rate in the massless limit. 

\bibliographystyle{JHEP-mod} 
\bibliography{references}

\providecommand{\href}[2]{#2}\begingroup\raggedright\begin{thebibliography}{10}

\bibitem{ATLAS:2012yve}
{\bf ATLAS}, G.~Aad {\it et.al.}, {\it {Observation of a new particle in the search for the Standard Model Higgs boson with the ATLAS detector at the LHC}},  Phys. Lett. B {\bf 716} (2012) 1--29, [\href{https://arxiv.org/abs/1207.7214}{{\tt arXiv:1207.7214}}].

\bibitem{CMS:2012qbp}
{\bf CMS}, S.~Chatrchyan {\it et.al.}, {\it {Observation of a New Boson at a Mass of 125 GeV with the CMS Experiment at the LHC}},  Phys. Lett. B {\bf 716} (2012) 30--61, [\href{https://arxiv.org/abs/1207.7235}{{\tt arXiv:1207.7235}}].

\bibitem{CMS:2013btf}
{\bf CMS}, S.~Chatrchyan {\it et.al.}, {\it {Observation of a New Boson with Mass Near 125 GeV in $pp$ Collisions at $\sqrt{s}$ = 7 and 8 TeV}},  JHEP {\bf 06} (2013) 081, [\href{https://arxiv.org/abs/1303.4571}{{\tt arXiv:1303.4571}}].

\bibitem{ATLAS:2022vkf}
{\bf ATLAS}, G.~Aad {\it et.al.}, {\it {A detailed map of Higgs boson interactions by the ATLAS experiment ten years after the discovery}},  Nature {\bf 607} (2022), no.~7917 52--59, [\href{https://arxiv.org/abs/2207.00092}{{\tt arXiv:2207.00092}}]. [Erratum: Nature 612, E24 (2022)].

\bibitem{CMS:2022dwd}
{\bf CMS}, A.~Tumasyan {\it et.al.}, {\it {A portrait of the Higgs boson by the CMS experiment ten years after the discovery.}},  Nature {\bf 607} (2022), no.~7917 60--68, [\href{https://arxiv.org/abs/2207.00043}{{\tt arXiv:2207.00043}}]. [Erratum: Nature 623, (2023)].

\bibitem{ATLAS:2015egz}
{\bf ATLAS}, G.~Aad {\it et.al.}, {\it {Measurements of the Higgs boson production and decay rates and coupling strengths using pp collision data at $\sqrt{s}=7$ and 8 TeV in the ATLAS experiment}},  Eur. Phys. J. C {\bf 76} (2016), no.~1 6, [\href{https://arxiv.org/abs/1507.04548}{{\tt arXiv:1507.04548}}].

\bibitem{CMS:2014fzn}
{\bf CMS}, V.~Khachatryan {\it et.al.}, {\it {Precise determination of the mass of the Higgs boson and tests of compatibility of its couplings with the standard model predictions using proton collisions at 7 and 8 $\,\text {TeV}$}},  Eur. Phys. J. C {\bf 75} (2015), no.~5 212, [\href{https://arxiv.org/abs/1412.8662}{{\tt arXiv:1412.8662}}].

\bibitem{CMS:2012vby}
{\bf CMS}, S.~Chatrchyan {\it et.al.}, {\it {Study of the Mass and Spin-Parity of the Higgs Boson Candidate Via Its Decays to Z Boson Pairs}},  Phys. Rev. Lett. {\bf 110} (2013), no.~8 081803, [\href{https://arxiv.org/abs/1212.6639}{{\tt arXiv:1212.6639}}].

\bibitem{ATLAS:2013xga}
{\bf ATLAS}, G.~Aad {\it et.al.}, {\it {Evidence for the spin-0 nature of the Higgs boson using ATLAS data}},  Phys. Lett. B {\bf 726} (2013) 120--144, [\href{https://arxiv.org/abs/1307.1432}{{\tt arXiv:1307.1432}}].

\bibitem{CMS:2014nkk}
{\bf CMS}, V.~Khachatryan {\it et.al.}, {\it {Constraints on the spin-parity and anomalous HVV couplings of the Higgs boson in proton collisions at 7 and 8 TeV}},  Phys. Rev. D {\bf 92} (2015), no.~1 012004, [\href{https://arxiv.org/abs/1411.3441}{{\tt arXiv:1411.3441}}].

\bibitem{CMS:2017dib}
{\bf CMS}, A.~M. Sirunyan {\it et.al.}, {\it {Measurements of properties of the Higgs boson decaying into the four-lepton final state in pp collisions at $ \sqrt{s}=13 $ TeV}},  JHEP {\bf 11} (2017) 047, [\href{https://arxiv.org/abs/1706.09936}{{\tt arXiv:1706.09936}}].

\bibitem{Super-Kamiokande:1998kpq}
{\bf Super-Kamiokande}, Y.~Fukuda {\it et.al.}, {\it {Evidence for oscillation of atmospheric neutrinos}},  Phys. Rev. Lett. {\bf 81} (1998) 1562--1567, [\href{https://arxiv.org/abs/hep-ex/9807003}{{\tt hep-ex/9807003}}].

\bibitem{SNO:2002tuh}
{\bf SNO}, Q.~R. Ahmad {\it et.al.}, {\it {Direct evidence for neutrino flavor transformation from neutral current interactions in the Sudbury Neutrino Observatory}},  Phys. Rev. Lett. {\bf 89} (2002) 011301, [\href{https://arxiv.org/abs/nucl-ex/0204008}{{\tt nucl-ex/0204008}}].

\bibitem{Arganda:2004bz}
E.~Arganda, A.~M. Curiel, M.~J. Herrero, and D.~Temes, {\it {Lepton flavor violating Higgs boson decays from massive seesaw neutrinos}},  Phys. Rev. D {\bf 71} (2005) 035011, [\href{https://arxiv.org/abs/hep-ph/0407302}{{\tt hep-ph/0407302}}].

\bibitem{Diaz-Cruz:1999sns}
J.~L. Diaz-Cruz and J.~J. Toscano, {\it {Lepton flavor violating decays of Higgs bosons beyond the standard model}},  Phys. Rev. D {\bf 62} (2000) 116005, [\href{https://arxiv.org/abs/hep-ph/9910233}{{\tt hep-ph/9910233}}].

\bibitem{Arhrib:2012ax}
A.~Arhrib, Y.~Cheng, and O.~C.~W. Kong, {\it {Comprehensive analysis on lepton flavor violating Higgs boson to $\mu^\mp \tau^\pm$ decay in supersymmetry without $R$ parity}},  Phys. Rev. D {\bf 87} (2013), no.~1 015025, [\href{https://arxiv.org/abs/1210.8241}{{\tt arXiv:1210.8241}}].

\bibitem{Goudelis:2011un}
A.~Goudelis, O.~Lebedev, and J.-h. Park, {\it {Higgs-induced lepton flavor violation}},  Phys. Lett. B {\bf 707} (2012) 369--374, [\href{https://arxiv.org/abs/1111.1715}{{\tt arXiv:1111.1715}}].

\bibitem{Pilaftsis:1992st}
A.~Pilaftsis, {\it {Lepton flavor nonconservation in H0 decays}},  Phys. Lett. B {\bf 285} (1992) 68--74.

\bibitem{Ishimori:2010au}
H.~Ishimori, T.~Kobayashi, H.~Ohki, Y.~Shimizu, H.~Okada, and M.~Tanimoto, {\it {Non-Abelian Discrete Symmetries in Particle Physics}},  Prog. Theor. Phys. Suppl. {\bf 183} (2010) 1--163, [\href{https://arxiv.org/abs/1003.3552}{{\tt arXiv:1003.3552}}].

\bibitem{Bjorken:1977vt}
J.~D. Bjorken and S.~Weinberg, {\it {A Mechanism for Nonconservation of Muon Number}},  Phys. Rev. Lett. {\bf 38} (1977) 622.

\bibitem{Branco:2011iw}
G.~C. Branco, P.~M. Ferreira, L.~Lavoura, M.~N. Rebelo, M.~Sher, and J.~P. Silva, {\it {Theory and phenomenology of two-Higgs-doublet models}},  Phys. Rept. {\bf 516} (2012) 1--102, [\href{https://arxiv.org/abs/1106.0034}{{\tt arXiv:1106.0034}}].

\bibitem{Agashe:2009di}
K.~Agashe and R.~Contino, {\it {Composite Higgs-Mediated FCNC}},  Phys. Rev. D {\bf 80} (2009) 075016, [\href{https://arxiv.org/abs/0906.1542}{{\tt arXiv:0906.1542}}].

\bibitem{Azatov:2009na}
A.~Azatov, M.~Toharia, and L.~Zhu, {\it {Higgs Mediated FCNC's in Warped Extra Dimensions}},  Phys. Rev. D {\bf 80} (2009) 035016, [\href{https://arxiv.org/abs/0906.1990}{{\tt arXiv:0906.1990}}].

\bibitem{Lami:2016mjf}
A.~Lami and P.~Roig, {\it {$H\to \ell\ell'$ in the simplest little Higgs model}},  Phys. Rev. D {\bf 94} (2016), no.~5 056001, [\href{https://arxiv.org/abs/1603.09663}{{\tt arXiv:1603.09663}}].

\bibitem{Han:2000jz}
T.~Han and D.~Marfatia, {\it $\mathit{h}\ensuremath{\rightarrow}\ensuremath{\mu}\ensuremath{\tau}$ at hadron colliders},  Phys. Rev. Lett. {\bf 86} (Feb, 2001) 1442--1445.

\bibitem{Brignole:2003iv}
A.~Brignole and A.~Rossi, {\it {Lepton flavor violating decays of supersymmetric Higgs bosons}},  Phys. Lett. B {\bf 566} (2003) 217--225, [\href{https://arxiv.org/abs/hep-ph/0304081}{{\tt hep-ph/0304081}}].

\bibitem{CMS:2023pte}
{\bf CMS}, A.~Hayrapetyan {\it et.al.}, {\it {Search for the lepton-flavor violating decay of the Higgs boson and additional Higgs bosons in the e$\mu$ final state in proton-proton collisions at $\sqrt{s}$ = 13 TeV}},  Phys. Rev. D {\bf 108} (2023), no.~7 072004, [\href{https://arxiv.org/abs/2305.18106}{{\tt arXiv:2305.18106}}].

\bibitem{CMS:2021rsq}
{\bf CMS}, A.~M. Sirunyan {\it et.al.}, {\it {Search for lepton-flavor violating decays of the Higgs boson in the $\mu\tau$ and e$\tau$ final states in proton-proton collisions at $\sqrt{s}$ = 13 TeV}},  Phys. Rev. D {\bf 104} (2021), no.~3 032013, [\href{https://arxiv.org/abs/2105.03007}{{\tt arXiv:2105.03007}}].

\bibitem{ATLAS:2019old}
{\bf ATLAS}, G.~Aad {\it et.al.}, {\it {Search for the Higgs boson decays $H \to ee$ and $H \to e\mu$ in $pp$ collisions at $\sqrt{s} = 13$ TeV with the ATLAS detector}},  Phys. Lett. B {\bf 801} (2020) 135148, [\href{https://arxiv.org/abs/1909.10235}{{\tt arXiv:1909.10235}}].

\bibitem{Ibarra:2021xyk}
A.~Ibarra, M.~Mar\'\i{}n, and P.~Roig, {\it {Flavor violating muon decay into an electron and a light gauge boson}},  Phys. Lett. B {\bf 827} (2022) 136933, [\href{https://arxiv.org/abs/2110.03737}{{\tt arXiv:2110.03737}}].

\bibitem{Marin:2022wfk}
M.~Mar\'\i{}n, {\em {\href{https://www.fis.cinvestav.mx/~proig/Tesis_Marcela.pdf}{Effective field theories in lepton flavor violating processes}}}.
\newblock PhD thesis, Cinvestav, 5, 2022.

\bibitem{Navas:2024prd}
S.~Navas {\it et.al.}, {\it Particle data group},  To be published in Physical Review D {\bf 110} (2024), no.~3 030001.

\end{thebibliography}\endgroup

\end{document}